\documentclass[aps,prb,preprint,superscriptaddress,floatfix,nobibnotes]{revtex4-1}

\usepackage{amsmath}
\usepackage{amssymb}
\usepackage{graphicx}
\usepackage[separate-uncertainty = true,multi-part-units=brackets]{siunitx}
\usepackage{hyperref}
\usepackage{array}
\usepackage{xr}
\usepackage[utf8]{inputenc}
\usepackage[english]{babel}
\usepackage{comment}
\newcommand{\lin}[1]{\mathrm{#1}}

\newcommand{\sigmaF}{\sigma_\lin{FL}^\lin{F}}

\newcommand{\thetaSH}{\theta_\lin{SH}}
\newcommand{\sigmaSH}{\sigma_\lin{SH}}

\newcommand{\sigmaFL}{\sigma_\lin{FL}} %field-like torque conductivity
\newcommand{\sigmaDL}{\sigma_\lin{DL}} %damping-like torque conductivity
\newcommand{\Meff}{M_\lin{eff}}
\newcommand{\Ms}{M_\lin{s}}

\newcommand{\dPt}{d_\lin{Pt}}
\newcommand{\Py}{Ni$_{80}$Fe$_{20}$}
\newcommand{\ls}{\lambda_\lin{s}} %spin diffusion length
\newcommand{\Gud}{G^{\uparrow\downarrow}} %spin mixing conductance

\DeclareSIUnit\torr{Torr}
\DeclareSIUnit\sq{\ensuremath{\Box}}

\begin{document}

%\preprint{}

\title{Determination of spin Hall effect and spin diffusion length of Pt from self-consistent fitting of damping enhancement and inverse spin-orbit torque measurements}

\author{Andrew J. Berger}
\affiliation{Quantum Electromagnetics Division, National Institute of Standards and Technology, Boulder, CO 80305, U.S.A.}
\thanks{Contribution of the National Institute of Standards and Technology; not subject to copyright.}

\author{Eric R. J. Edwards}
\affiliation{Quantum Electromagnetics Division, National Institute of Standards and Technology, Boulder, CO 80305, U.S.A.}

\author{Hans T. Nembach}
\affiliation{Quantum Electromagnetics Division, National Institute of Standards and Technology, Boulder, CO 80305, U.S.A.}

%\author{Justin M. Shaw}
%\affiliation{Quantum Electromagnetics Division, National Institute of Standards and Technology, Boulder, CO 80305, U.S.A.}

\author{Olof Karis}
\affiliation{Department of Physics and Astronomy, Uppsala University,  Box 530, 751 20 Uppsala}

\author{Mathias Weiler}
\affiliation{Walther-Mei{\ss}ner-Institut, Bayerische Akademie der Wissenschaften, Garching, Germany}
\affiliation{Physik-Department, Technische Universität München, Garching, Germany}

\author{T. J. Silva}
\email[]{thomas.silva@nist.gov}
\affiliation{Quantum Electromagnetics Division, National Institute of Standards and Technology, Boulder, CO 80305, U.S.A.}

\date{\today}

\begin{abstract}
Understanding the evolution of spin-orbit torque (SOT) with increasing heavy-metal thickness in ferromagnet/normal metal (FM/NM) bilayers is critical for the development of magnetic memory based on SOT. However, several experiments have revealed an apparent discrepancy between damping enhancement and damping-like SOT regarding their dependence on NM thickness. Here, using linewidth and phase-resolved amplitude analysis of vector network analyzer ferromagnetic resonance (VNA-FMR) measurements, we simultaneously extract damping enhancement and both field-like and damping-like inverse SOT in \Py/Pt bilayers as a function of Pt thickness. By enforcing an interpretation of the data which satisfies Onsager reciprocity, we find that both the damping enhancement and damping-like inverse SOT can be described by a single spin diffusion length ($\approx$ \SI{4}{\nm}), and that we can separate the spin pumping and spin memory loss contributions to the total damping. This analysis indicates that less than 40\% of the angular momentum pumped by FMR through the \Py/Pt interface is transported as spin current into the Pt. On account of the spin memory loss and corresponding reduction in total spin current available for spin-charge transduction in the Pt, we determine the Pt spin Hall conductivity ($\sigmaSH = $ \SI{2.36(4)e6}{\ohm^{-1}\m^{-1}}) and bulk spin Hall angle ($\thetaSH = $ \SI{0.387(8)}{}) to be larger than commonly-cited values. These results suggest that Pt can be an extremely useful source of SOT if the FM/NM interface can be engineered to minimize spin loss. Lastly, we find that self-consistent fitting of the damping and SOT data is best achieved by a model with Elliott-Yafet spin relaxation and extrinsic inverse spin Hall effect, such that both the spin diffusion length and spin Hall conductivity are proportional to the Pt charge conductivity.
\end{abstract}

\maketitle

\section{Introduction}
The use of nonmagnetic metals with strong spin-orbit coupling (SOC) to generate pure spin currents via spin-orbit effects is currently an area of intense focus, driven largely by the promise of efficient electrically-controllable magnetic memory. For this application, the spin current or spin accumulation generated by SOC in a non-magnetic layer can be used to exert a torque on an adjacent ferromagnetic (FM) layer---so called spin-orbit torque (SOT)---in order to excite magnetization dynamics or cause switching. Central to this field of study is proper characterization of the spin-to-charge conversion that occurs in heavy metal films such as Pt, Ta, W, and Au. There are many techniques for measuring this conversion, including ferromagnetic resonance (FMR) spin pumping \cite{saitoh_conversion_2006}, non-local spin valves \cite{valenzuela_direct_2006, sagasta_tuning_2016}, thermal spin injection via the spin Seebeck effect \cite{qu_self-consistent_2014}, spin Hall magnetoresistance \cite{nakayama_spin_2013}, spin torque FMR \cite{liu_spin-torque_2011}, and harmonic analysis of Hall effect voltage measurements \cite{garello_symmetry_2013}. Several groups, using various techniques \cite{nakayama_geometry_2012, feng_spin_2012, rojas-sanchez_spin_2014, nan_comparison_2015, conca_lack_2017}, have uncovered a discrepancy when comparing the excess damping and the spin-to-charge conversion by inverse spin Hall effect (iSHE) contributed by the normal metal (NM) layer. Specifically, the FM damping exhibits a steep increase with the introduction of only a very thin ($<$ \SI{2}{\nm}) NM film \cite{boone_spin-scattering_2015, azzawi_evolution_2016, caminale_spin_2016}. Meanwhile, the measured SOT, characterized by either spin-to-charge conversion via DC iSHE or harmonic Hall technique, develops over a much longer length length scale \cite{azevedo_spin_2011, nguyen_spin_2016}. Magneto-optical measurements also demonstrate an interfacial spin accumulation in Pt due to SHE with a  spin diffusion length of about \SI{10}{\nm} \cite{stamm_magneto-optical_2017}. 

Spin memory loss (SML) \cite{kurt_spin-memory_2002, rojas-sanchez_spin_2014} and proximity-induced magnetic moments at the FM/NM interface \cite{caminale_spin_2016} have been invoked to explain the large damping enhancement caused by thin NM films even when the NM thickness is less than its spin diffusion length. In this model, spin loss at the FM/NM interface acts as an additional parallel spin relaxation pathway to that of spin pumping and diffusion into the Pt bulk. From damping measurements alone, the relative contributions of these mechanisms is not resolvable. In this work, we show that a self-consistent fit of Gilbert damping and damping-like iSOT versus Pt thickness---where both sets of data are described by the same spin diffusion length $\ls$---makes it possible to separate these sources of damping. Furthermore, this data analysis methodology allows for unambiguous determination of the spin-mixing conductance $\Gud$ at the FM/NM interface. We therefore can ascertain the spin Hall conductivity (or spin Hall angle) without having to refer to spin transport parameters $\Gud$ and $\ls$ determined from measurements performed on dissimilar samples or theoretical idealized values.  For our samples of Pt deposited on \Py (or Permalloy, Py), only \SI{37(6)}{}\% of the total damping enhancement from the Pt film is attributable to spin pumping into the Pt layer when $\dPt \gg \ls$. 

%It is a necessary logical conclusion that with less spin current driven into the NM (on account of SML), the spin-to-charge conversion efficiency must increase in order to give rise to the observed charge signals. Taking SML into account, we find a spin Hall conductivity of $\sigmaSH = $ \SI{2.36(4)e6}{\ohm^{-1}\meter^{-1}}, corresponding to a bulk spin Hall angle of $\thetaSH \equiv \sigmaSH/\sigma_\lin{Pt}^\lin{bulk} = 0.387 \pm 0.008$. While these values for $\sigmaSH$ and $\thetaSH$ are larger than commonly reported values \cite{hoffmann_spin_2013}, the phenomenological value obtained for the damping-like SOT conductivity, $\sigmaDL \approx $ \SI{5e5}{\ohm^{-1}\meter^{-1}}, is comparable to what has been seen before \cite{garello_symmetry_2013, pai_dependence_2015, nguyen_spin_2016}.

%nguyen_spin_2016: \xi_DL = \sigmaDL = 3e5 ohm^-1 m^-1
%garello_symmetry_2013: \sigmaDL = 5.8e5 ohm^-1 m^-1
%pai: \xi_DL (efficiency) = 0.12
% \xi_DL*sigma_Pt = 4.99e5 ohm^-1 m^-1

%\section{Methods}
\section{Experimental Technique}
The data presented in this work are based on the spectroscopic and complex amplitude information encoded in VNA-FMR spectra, which yield a measure of the damping and SOT, respectively. FMR damping extracted from a spectral linewidth analysis \cite{kalarickal_ferromagnetic_2006} has been used extensively to study the damping enhancement due to the spin pumping effect into an NM adjacent to the FM layer \cite{tserkovnyak_enhanced_2002, mizukami_effect_2002, heinrich_role_2003, schoen_magnetic_2017-1}. If such spectra are measured inductively with phase-sensitive VNA-FMR, it is also possible to analyze the phase and amplitude information of those spectra to quantitatively extract the field-like (FL) and damping-like (DL) SOT conductivities, as we have previously described \cite{berger_inductive_2018}. These conductivities, $\sigmaFL^\lin{SOT}$ and $\sigmaDL^\lin{SOT}$, relate the AC charge currents produced in the NM layer via iSHE or inverse Rashba-Edelstein effect (iREE) in response to driven magnetization dynamics in the FM layer. Direct coupling to the magnetization dynamics via Faraday's law also drives AC charge currents in the NM layer, quantified by $\sigmaFL^\lin{F}$. The superposition of these charge currents presents a complex inductive load to the microwave coplanar waveguide (CPW) used in VNA-FMR measurements, altering the amplitude and phase of the transmitted microwave signal. By Onsager reciprocity, $\sigmaFL^\lin{SOT}$ and $\sigmaDL^\lin{SOT}$ measured inductively via inverse spin-charge conversion processes are equivalent to the spin torque efficiency per unit applied electric field used by Nguyen et al. in Ref. \citenum{nguyen_spin_2016} to describe the forward SOT process \cite{berger_inductive_2018}.

\subsection{Samples}
To study the Pt-thickness dependence of damping and damping-like iSOT, we prepared two sample sets, with sputter-deposited metal multilayers consisting of substrate\slash{}Ta(1.5)\slash{}Py($d_\lin{Py}$)\slash{}Pt($d_\lin{Pt}$)\slash{}Ta(3), where thicknesses are indicated in nanometers and are calibrated with X-ray reflectivity measurements. In the first sample set, the thickness $d_\lin{Py}$ was varied from \SI{1.5}{\nm} to \SI{10}{\nm} while $d_\lin{Pt} = $ \SI{6}{\nm} was fixed. In the second set, the thickness $d_\lin{Pt}$ was varied from \SI{2}{nm} to \SI{20}{\nm} with fixed $d_\lin{Py} = $ \SI{3.5}{\nm}. For each sample, an identical control sample was prepared, where Pt is substituted with Cu. The Cu thicknesses were chosen to match the sheet resistance of the corresponding Pt layer, so as to control for Faraday effect induced currents in the NM layer.

\section{Results and Discussion}
\subsection{Py thickness series}
From the Py thickness series we focus on three quantities: (1) the FM contribution to the sample inductance ($L_\lin{FM}$, as in Ref. \citenum{berger_inductive_2018}), (2) the effective magnetization $\Meff$, and (3) the Gilbert damping parameter $\alpha$. From $L_\lin{FM}$ as a function of Py thickness (Fig. \ref{fig:LFM_dPy}), we are able to extract the dead layer thickness, and therefore determine the effective magnetic thickness of the FM layer. From $\Meff$, we are able to determine the saturation magnetization $\Ms$ (Fig. \ref{fig:Meff_dPy}). Lastly, from the Gilbert damping as a function of Py thickness, we can separate the intrinsic and interfacial contributions to $\alpha$ (Fig. \ref{fig:alpha_dPy}). This is a critical first step to determine the spin pumping and SML contributions to the total damping.

\subsubsection{Ferromagnetic dead layer measurement}
\begin{figure}[hb]
	\centering
	\includegraphics[width=0.5\linewidth]{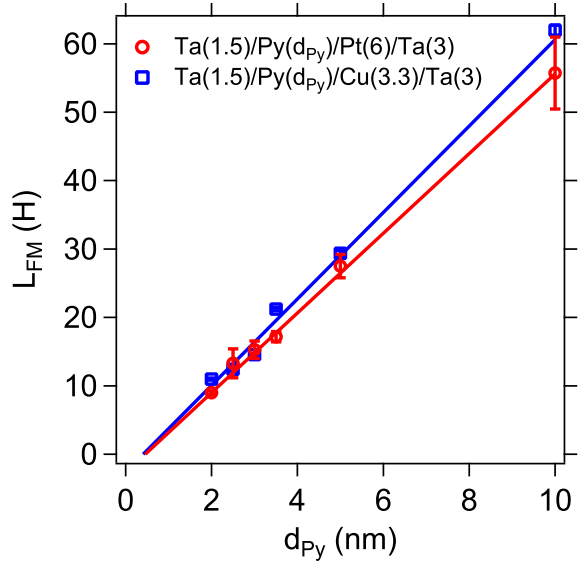}
	\caption{Py-thickness dependent zero-frequency inductance for both Py/Pt and Py/Cu control samples.}
\label{fig:LFM_dPy}
\end{figure}

In inductive VNA-FMR measurements, the FM layer contributes a frequency-independent inductance to the $S_{21}$ measurement according to \cite{berger_inductive_2018, silva_characterization_2016}:

\begin{equation}
L_\lin{FM} = \frac{\mu_0 l d_\lin{FM}}{4 W_\lin{wg}} \eta^2(z, W_\lin{wg})
\end{equation}

\noindent where $\mu_0$ is the permeability of free space, $l$ is the sample length along the CPW signal propagation direction, $d_\lin{FM}$ is the deposited FM thickness, $W_\lin{wg}$ is the CPW signal line width, and $\eta(z,W_\lin{wg}) \equiv (2/\pi)\arctan(W_\lin{wg}/2z)$ is the spacing loss, ranging from 0 to 1, due to a finite distance $z$ between sample and CPW. When plotted vs. $d_\lin{FM}$, the $L_\lin{FM} = 0$ intercept indicates the magnetic dead layer thickness. From the data in Fig. \ref{fig:LFM_dPy}, we find $d_\lin{dead} = $ \SI{0.5(1)}{\nm} for Py/Pt samples. Also shown are the data for Py/Cu control samples, which exhibit a similar dead layer thickness of \SI{0.41(4)}{\nm}, suggesting that the Py dead layer exists primarily at the Ta/Py interface.

%From ``Py thickness series plots for PRL SI.pxp''

\subsubsection{Determination of $\Ms$}
The effective magnetization $\Meff$ as a function of applied microwave frequency is extracted from the FMR spectral fits and the Kittel FMR condition for magnetization oriented out of the film plane \cite{kittel_introduction_2004}:

\begin{equation}
H_\lin{res} = \frac{\omega}{\gamma \mu_0} + \Meff
\end{equation}

\noindent where $H_\lin{res}$ is the center field of the resonant absorption line, $\omega$ is the applied microwave frequency, and $\gamma = g \mu_\lin{B}/\hbar$ is the gyromagnetic ratio. Assuming the Py has no bulk anisotropy, $\Meff$ is determined by the saturation magnetization $\Ms$ of the material, and the interfacial anisotropy energy $K_\lin{int}$ according to Ref. \citenum{schoen_magnetic_2017}:

\begin{equation}
\mu_0 \Meff = \mu_0 \Ms - \frac{2 K_\lin{int}}{\Ms} \left( \frac{1}{d_\lin{FM} - d_\lin{dead}} \right)
\end{equation}

\begin{figure}
	\centering
	\includegraphics[width=0.5\linewidth]{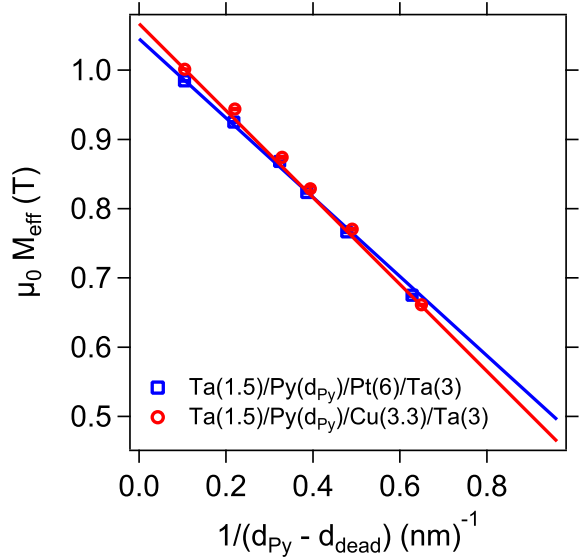}
	\caption{$\Meff$ vs. inverse effective FM thickness $(d_\lin{Py} - d_\lin{dead})$ for Py($d_\lin{Py})$/Pt(6) and Py($d_\lin{Py})$/Cu(3.3). Dead layer thickness is determined from Fig. \ref{fig:LFM_dPy}.}
\label{fig:Meff_dPy}
\end{figure}

\noindent Therefore, a linear fit of $\Meff$ vs. inverse effective FM thickness (Fig. \ref{fig:Meff_dPy}) provides a measurement of the saturation magnetization $\Ms$. We find $\mu_0 M_\lin{s} = $ \SI{1.0671(1)}{\tesla}, comparable to previous findings \cite{schoen_magnetic_2017}. Similarly, for Py/Cu we find $\mu_0 M_\lin{s} = $ \SI{1.0453(4)}{\tesla}.

\subsubsection{Determination of intrinsic Gilbert damping constant}
\begin{figure}
	\centering
	\includegraphics[width=0.5\linewidth]{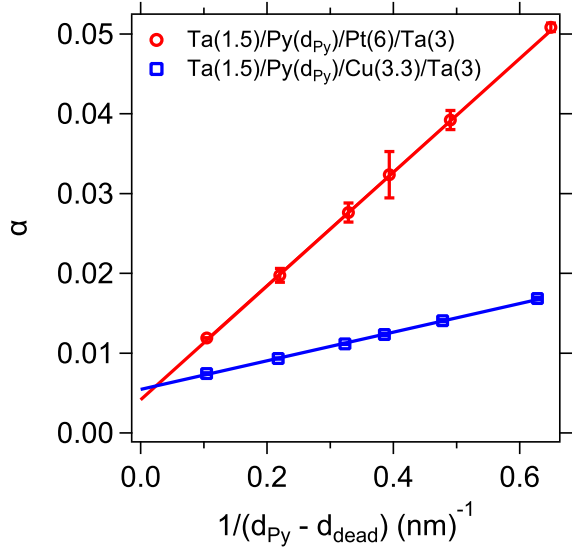}
	\caption{Total Gilbert damping vs. inverse effective FM thickness $(d_\lin{Py} - d_\lin{dead})$ for Py($d_\lin{Py}$)/Pt(6) (circles) and Py($d_\lin{Py}$)/Cu(3.3) (squares). Dead layer thickness is determined from Fig. \ref{fig:LFM_dPy}.}
\label{fig:alpha_dPy}
\end{figure}

The total Gilbert damping due to intrinsic and interfacial contributions can be described by:

\begin{equation}
\alpha = \alpha_\lin{int} + \Gud_\lin{eff}\left(\frac{\gamma \hbar^2}{2 \Ms d_\lin{FM} e^2}\right)
\label{eq:alphaTot}
\end{equation}

\noindent where $\gamma = g \mu_\lin{B}/\hbar$ is the gyromagnetic ratio, $g$ is the spectroscopic g factor, $\mu_\lin{B}$ is the Bohr magneton, $\hbar$ is Planck's constant divided by $2 \pi$, and $e$ is the electron charge. $M_\lin{s}$ and $d_\lin{FM}$ for the Py layer are determined as described above. For the thin FM layers studied here, we can ignore the contribution from radiative damping \cite{schoen_radiative_2015}. When plotted vs. $1/(d_\lin{Py} - d_\lin{dead})$, we can extract $\alpha_\lin{int}$ as the infinite-thickness limit of the measured damping. We calculate the intercept of the data in Fig. \ref{fig:alpha_dPy} using linear regression in order to fix $\alpha_\lin{int} = 0.0041 \pm 0.0001$, in good agreement with a previous systematic study of damping in magnetic alloys \cite{schoen_magnetic_2017}. 

\begin{figure}
	\centering
	\includegraphics[width=\linewidth]{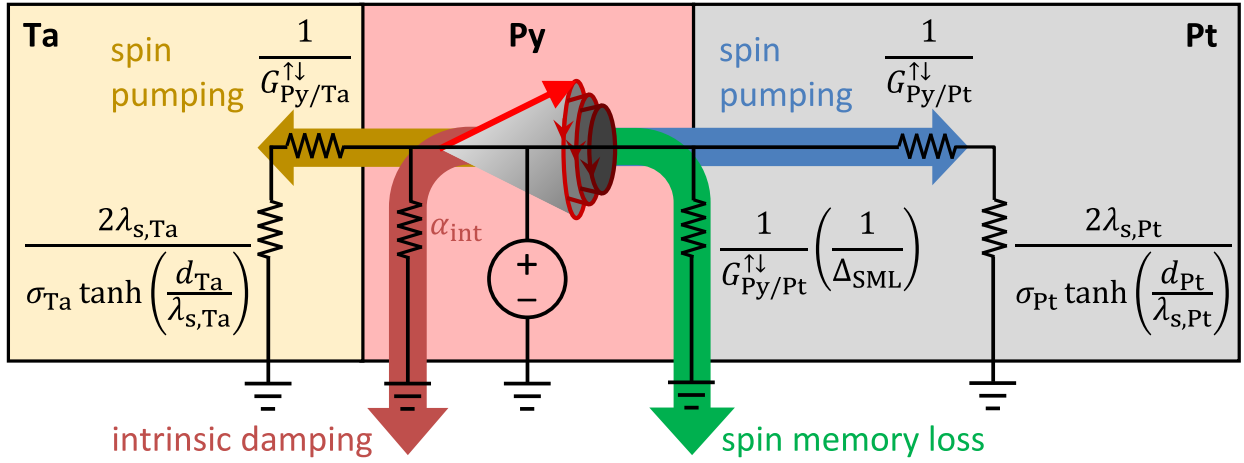}
	\caption{Circuit model for angular momentum flow sourced by FMR excitation in Ta/Py/Pt trilayer. Spin current is drawn into parallel resistance channels provided by spin pumping into the Ta seed and Pt spin sink layers, as well as spin memory loss.}
\label{fig:SpinPumping_CurrentDivider}
\end{figure}

For the interfacial contribution to the damping (second term in Eq. \eqref{eq:alphaTot}), the full model we use for the effective spin-mixing conductance $\Gud_\lin{eff}$ includes contributions from spin pumping into Pt via the spin-mixing conductance $\Gud_\lin{Py/Pt}$, spin pumping into the Ta seed layer via the spin-mixing conductance $\Gud_\lin{Py/Ta}$, and spin memory loss (SML). (In all instances where we invoke the spin-mixing conductance, it is to be understood that we are only considering the real part of said quantity).

\begin{widetext}
\begin{equation}
\Gud_\lin{eff} = \frac{\Gud_\lin{Py/Pt}}{1 + \dfrac{2 \lambda_\lin{s,Pt} \Gud_\lin{Py/Pt}}{\sigma_\lin{Pt}(d_\lin{Pt}) \tanh\left(\dfrac{d_\lin{Pt}}{\lambda_\lin{s,Pt}}\right)}} + \frac{\Gud_\lin{Py/Ta}}{1 + \dfrac{2 \lambda_\lin{s,Ta} \Gud_\lin{Py/Ta}}{\sigma_\lin{Ta}(d_\lin{Ta}) \tanh\left(\dfrac{d_\lin{Ta}}{\lambda_\lin{s,Ta}}\right)}} + \Gud_\lin{Py/Pt} \Delta_\lin{SML}
\label{eq:Geff}
\end{equation}
\end{widetext}

\noindent This model is depicted as a network of series and parallel conductance channels for the flow of angular momentum, treating FMR as an angular momentum potential source, as depicted in Fig. \ref{fig:SpinPumping_CurrentDivider} (also see Ref. \citenum{roy_estimating_2017}). The first two terms of Eq. \eqref{eq:Geff} represent spin pumping into the Pt and Ta layers, respectively. Within those layers, spin is pumped through series resistances set by the interfacial spin-mixing conductance, and thickness-dependent spin resistance (which accounts for the exponential spin accumulation profile in the NM layer, as a solution to the spin diffusion equation, subject to the boundary condition that no spin current can flow through the distant interface). The final term represents a spin memory loss channel, where the phenomenological parameter $\Delta_\lin{SML}$ can be arbitrarily large. By multiplying Eq. \eqref{eq:Geff} by the bracketed term in Eq. \eqref{eq:alphaTot}, conductances are converted to the unitless damping parameters $\alpha_\lin{sp,Pt(Ta)}$ (due to spin pumping into Pt (or Ta)) and $\alpha_\lin{SML}$ (due to spin memory loss). 

Taken together, Eqs. \eqref{eq:alphaTot} and \eqref{eq:Geff} describe both the NM and FM thickness dependencies of the damping. As a part of our self-consistent fitting routine (described in Section \ref{sec:selfConsistent}, and using the previously determined value for $\alpha_\lin{int}$, we fit the Py thickness dependence of $\alpha$ simultaneously with the Pt thickness dependence (Fig. \ref{fig:sigmaSOTandAlpha}(b)), with $\Gud$ and $\Delta_\lin{SML}$ as fit parameters. The result of that simultaneous fit is shown in Fig. \ref{fig:alpha_dPy}. Also shown for comparison are damping data for the Py/Cu controls, which exhibit a drastically reduced spin pumping contribution (slope), and slightly increased intrinsic contribution ($\alpha_\lin{int} = 0.0054 \pm 0.0001$). 

\subsection{Pt thickness series}
\label{sec:PtThickness}
\begin{figure}
	\centering
	\includegraphics[width=\linewidth]{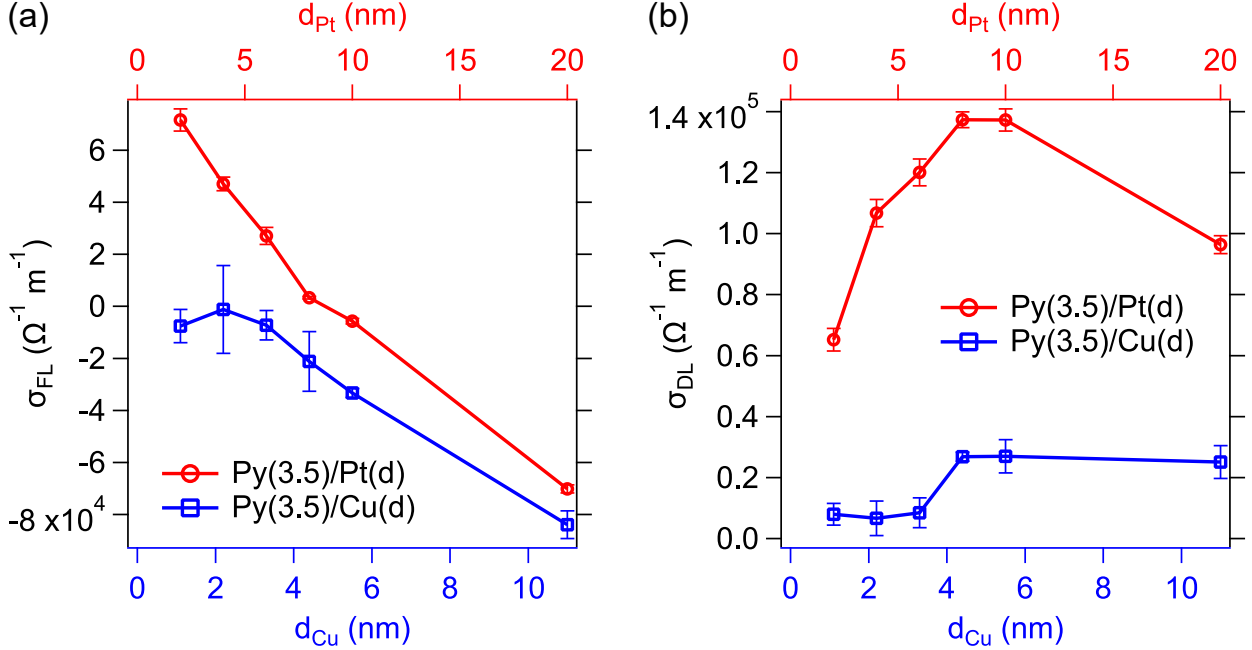}
	\caption{Measured quantities for FL and DL conductivities, for both Pt and Cu control samples, extracted from complex inductance analysis of VNA-FMR data \cite{berger_inductive_2018}. (a) $\sigmaFL$ as a function of either Pt (top axis) or Cu thickness (bottom axis). Linear dependence on NM thickness at large thicknesses indicates dominance of $\sigmaF$ term. (b) Same as (a), but for $\sigmaDL$. }
\label{fig:sigmaFL_DL}
%from ``Pt thickness series, Tom's correction results.pxp''
\end{figure}

\begin{figure}
	\centering
	\includegraphics[width=\linewidth]{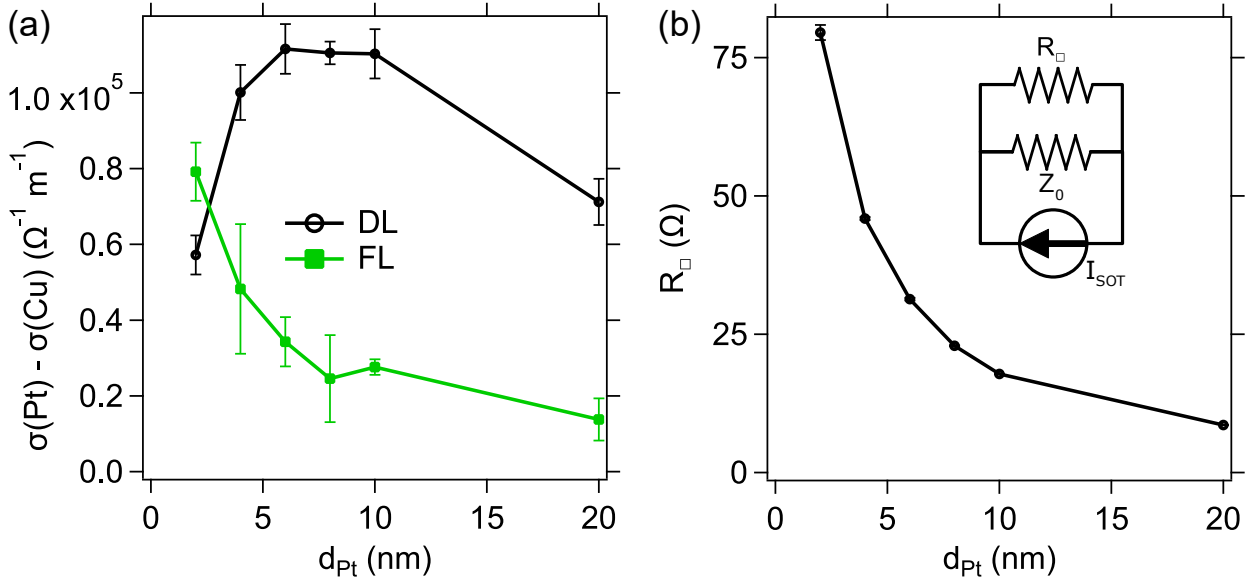}
	\caption{(a) FL and DL iSOT conductivities, after subtraction of Cu control samples. (b) Measured sheet resistance of metallic layers, as a function of Pt thickness. Inset: the sample sheet resistance acts as a parallel shunting path to the signal generating component of $I_\lin{SOT}$, which flows through the characteristic impedance $Z_0$ and $R_{\square}$.}
\label{fig:ShuntCorr}
%from ``Pt thickness series, Tom's correction results.pxp''
\end{figure}

For samples where the Pt thickness is varied, the measured values for $\sigmaFL^\lin{SOT}$ and $\sigmaDL^\lin{SOT}$---extracted from our quantitative VNA-FMR complex amplitude analysis \cite{berger_inductive_2018}---are shown as a function of NM thickness in Fig. \ref{fig:sigmaFL_DL}. Two corrections must be made to these values in order to extract the iSOT due to Pt. First, we subtract the values for $\sigmaFL$ and $\sigmaDL$ obtained from the Cu control samples (blue squares) from those of the Pt samples (red circles). Since we used Cu thicknesses to match the sheet resistance of the Pt samples, this removes the Faraday contribution. This subtraction also removes any FL or DL iSOT due to the Ta seed and capping layers. While we do not completely understand the Cu thickness dependence of $\sigmaDL$, the DL signal is essentially eliminated for Py in isolation, without seed or capping layers, which suggests details of the iSOT from cap and/or seed layers are responsible for the peculiar behavior (see discussion and measurements in Section \ref{sec:ControlSamples}). In Fig. \ref{fig:ShuntCorr}(a), $\sigmaFL$ and $\sigmaDL$ after Cu reference subtraction are plotted. 

Second, we correct for shunting effects of the iSOT currents. The data of Fig. \ref{fig:ShuntCorr}(a) attenuate as the Pt thickness is increased. This is attributed to the decreasing sheet resistance of the metallic stack, which effectively shunts the AC iSOT currents, therefore producing a weaker inductive response. This is functionally similar to the current divider effect observed in DC voltage iSHE spin pumping experiments \cite{nakayama_geometry_2012, feng_spin_2012, jiao_spin_2013, wang_scaling_2014}. However, in our AC iSOT experiments with the sample placed on a CPW with characteristic impedance of \SI{50}{\ohm}, the sample sheet resistance acts as a shunt path in parallel with the CPW characteristic impedance (inset of Fig. \ref{fig:ShuntCorr}(b)). We therefore multiply the $\sigmaFL$ and $\sigmaDL$ results of Fig. \ref{fig:ShuntCorr}(a) by the shunt factor $(1 + Z_0/R_{\square})$, where $R_{\square}$ is the measured sheet resistance of the multilayer stack (Fig. \ref{fig:ShuntCorr}(b)).

\begin{figure}
	\centering
	\includegraphics[width=\linewidth]{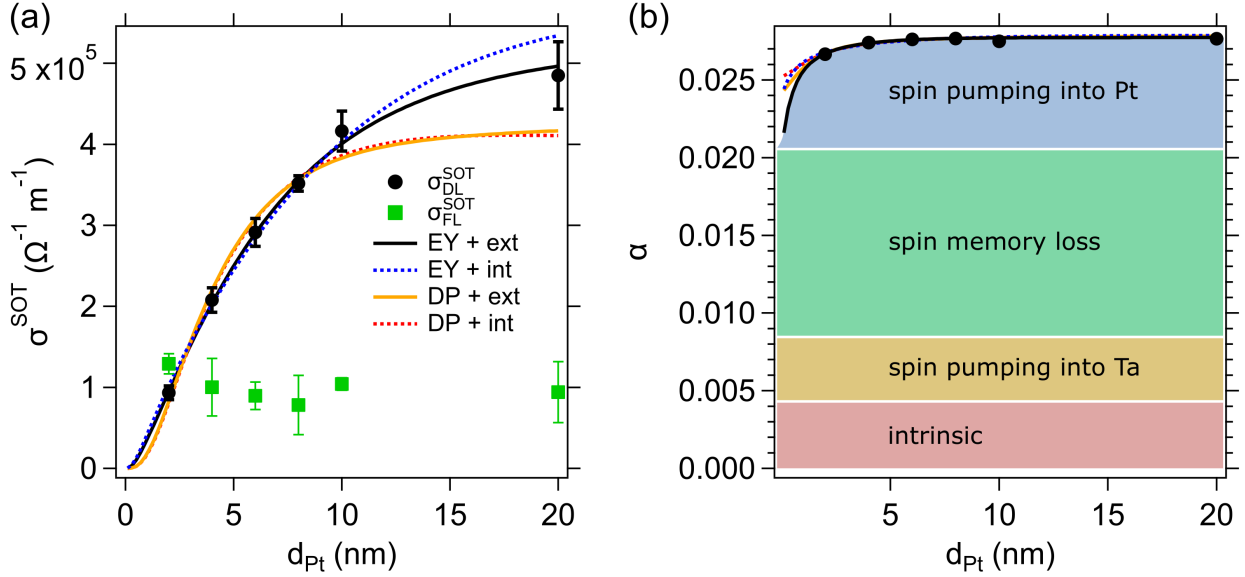}
	\caption{(a) Final values for $\sigmaDL^\lin{SOT}$ and $\sigmaFL^\lin{SOT}$ for Py(3.5)/Pt($\dPt$). The FL torque remains constant over the range of studied thicknesses, whereas the DL torque increases with a characteristic length scale. (b) Gilbert damping for the same sample series (error bars are smaller than symbols). Color coding indicates different contributions to the Gilbert damping. Both the SOT conductivity and damping are fit to four different models, where spin relaxation is either EY or DP, and the spin Hall effect arises from intrinsic (int) or extrinsic (ext) processes. In both cases EY + ext (black solid line) provides the best fit, as determined by a $\chi^2$ analysis.}
\label{fig:sigmaSOTandAlpha}
%from ``Pt thickness series, Tom's correction results.pxp''
%from ``Python fitting results, iterative SOT and damping.pxp''
%from ``Gilbert Damping vs Pt thickness, PtTS.pxp''
\end{figure}

After application of the shunting correction, the final results for $\sigmaFL^\lin{SOT}$ and $\sigmaDL^\lin{SOT}$ are presented in Fig. \ref{fig:sigmaSOTandAlpha}(a). These results are shown adjacent to the dependence of the measured Gilbert damping parameter $\alpha$ on Pt thickness in Fig. \ref{fig:sigmaSOTandAlpha}(b) to compare their evolution with $d_\lin{Pt}$. The DL conductivity increases monotonically with Pt thickness. Meanwhile, the FL conductivity remains more or less constant, consistent with the presumption of an interfacial source of spin-charge conversion such as iREE, where additional Pt beyond \SI{2}{\nm} does not increase the charge signal further. From Fig. \ref{fig:sigmaSOTandAlpha}(b), it is clear that if the enhanced damping (second term in Eq. \eqref{eq:alphaTot}) were ascribed entirely to spin pumping into the Pt, the length scale necessary to capture the rapid increase in $\alpha$ above the intrinsic value must be much shorter than the length scale over which $\sigmaDL$ is seen to increase in \ref{fig:sigmaSOTandAlpha}(a). In other words, using only the data for Pt-thickness dependence of damping (Fig. \ref{fig:sigmaSOTandAlpha}(b)), it is impossible to separate the different contributions to $\Gud_\lin{eff}$. Several other groups have observed this apparent discrepancy when comparing damping with DC voltages measured by iSHE \cite{nakayama_geometry_2012, feng_spin_2012, rojas-sanchez_spin_2014}. In this work, we are able to resolve the discrepancy through a self-consistent fit of both the damping data and $\sigmaDL^\lin{SOT}$ versus Pt thickness. 

%In the simplest model, $\Gud_\lin{eff} = \Gud/(1 + 2 \ls \Gud/(\sigma_\lin{NM}(d_\lin{NM} \tanh(d_\lin{NM}/\ls)))$, to account for spin diffusion through the thickness of the NM. In this case, the $\tanh(d_\lin{NM}/\ls)$ function is responsible for the thickness-dependence, and reaches 96\% of it's saturation value at $d_\lin{NM} = 2 \ls$.

Although we measure only a 3\% enhancement of damping as $\dPt$ increases from \SI{2}{\nm} to \SI{20}{\nm}, given the high signal-to-noise ratio of the damping data, it can be fit with Eq. \eqref{eq:alphaTot} and \eqref{eq:Geff} by use of the same spin diffusion length that describes the behavior of $\sigmaDL^\lin{SOT}$, as discussed in detail later. Because of the better dynamic range of the $\sigmaDL^\lin{SOT}$ data, we use it as the basis for establishing $\ls$ by fitting with a model provided by Haney et al. \cite{haney_current_2013}:

%for preprint, single-line equation
\begin{widetext}
\begin{equation}
\sigmaDL^\lin{SOT} = \sigmaSH \left\{\frac{(1 - e^{-d_\lin{NM}/\lambda_\lin{s}})^2}{(1 + e^{-2 d_\lin{NM}/\lambda_\lin{s}})} 
\frac{|\tilde{G}^{\uparrow\downarrow}|^2 + \mbox{Re}(\tilde{G}^{\uparrow\downarrow})\tanh^2\left(\dfrac{d_\lin{NM}}{\lambda_\lin{s}}\right)}
{|\tilde{G}^{\uparrow\downarrow}|^2 + 2 \mbox{Re}(\tilde{G}^{\uparrow\downarrow})\tanh^2\left(\dfrac{d_\lin{NM}}{\lambda_\lin{s}}\right) + \tanh^4\left(\dfrac{d_\lin{NM}}{\lambda_\lin{s}}\right)}\right\} \epsilon
\label{eq:StilesHaney}
\end{equation}
\end{widetext}

%for reprint, multline equation
\begin{comment}
\begin{multline}
\sigmaDL^\lin{SOT} = \sigmaSH \left\{\frac{(1 - e^{-d_\lin{NM}/\lambda_\lin{s}})^2}{(1 + e^{-2 d_\lin{NM}/\lambda_\lin{s}})}\right. \\
\left.*\frac{|\tilde{G}^{\uparrow\downarrow}|^2 + \mbox{Re}(\tilde{G}^{\uparrow\downarrow})\tanh^2\left(\dfrac{d_\lin{NM}}{\lambda_\lin{s}}\right)}
{|\tilde{G}^{\uparrow\downarrow}|^2 + 2 \mbox{Re}(\tilde{G}^{\uparrow\downarrow})\tanh^2\left(\dfrac{d_\lin{NM}}{\lambda_\lin{s}}\right) + \tanh^4\left(\dfrac{d_\lin{NM}}{\lambda_\lin{s}}\right)}\right\} \epsilon
\label{eq:StilesHaney}
\end{multline}
\end{comment}

\noindent where $\tilde{G}^{\uparrow\downarrow} = \Gud 2 \lambda_\lin{s} \tanh(d_\lin{NM}/\lambda_\lin{s})/\sigma$, $\sigma$ represents the NM charge conductivity, and $\epsilon \equiv \alpha_\lin{sp,Pt}/(\alpha_\lin{sp,Pt} + \alpha_\lin{SML})$ represents the fraction of spin current pumped out of the FM that is available for spin-charge conversion in the Pt layer, as determined by the the spin current divider model applied to the first and last terms of Eq. \eqref{eq:Geff}.

\subsection{Self-consistent fit routine of damping and SOT}
\label{sec:selfConsistent}
To perform the self-consistent fits of $\sigmaDL^\lin{SOT}$ and $\alpha$, an initial fit of $\sigmaDL^\lin{SOT}$ is performed to extract the Pt spin diffusion length $\lambda_\lin{s,Pt}$. This is then used as a fixed parameter in Eq. \eqref{eq:alphaTot} and \eqref{eq:Geff} when fitting $\alpha$. With this constraint on $\lambda_\lin{s,Pt}$, the Pt and Py thickness series (Figs. \ref{fig:sigmaSOTandAlpha}(b) and \ref{fig:alpha_dPy}, respectively) are fitted simultaneously with Eq. \eqref{eq:alphaTot} and \eqref{eq:Geff} to determine $\Gud_\lin{Py/Pt}$ and $\Delta_\lin{SML}$. These are then put back into Eq. \eqref{eq:StilesHaney} to re-fit $\sigmaDL^\lin{SOT}$ and extract refined values for $\sigmaSH$ and $\lambda_\lin{s,Pt}$. This process is iterated until the change in fit parameters is less than 0.01\%.

Our self-consistent data analysis is tantamount to enforcing Onsager reciprocity on the spin-to-charge interconversion processes of spin pumping and spin torque \cite{brataas_spin_2012}. If the enhanced damping of Fig. \ref{fig:sigmaSOTandAlpha}(b) were ascribed purely to spin pumping, it would imply that the Pt already draws a maximum amount of spin current from the precessing FM at thicknesses of only $\approx$ \SI{2}{\nm}. By contrast, a damping-like torque conductivity that continues to increase for thicknesses up to \SI{10}{\nm} (Fig. \ref{fig:sigmaSOTandAlpha}(a)) suggests that the Pt layer can continue to generate (or draw) increasingly larger spin current for thicknesses well beyond \SI{2}{\nm}. The use of unequal length scales to describe diffusive spin current flow due to spin pumping and spin-orbit torque generation would violate the reciprocity of spin-to-charge interconversion. 

Equations \eqref{eq:Geff} and \eqref{eq:StilesHaney} can be used either with an Elliott-Yafet (EY) \cite{elliott_theory_1954, yafet_conduction_1983} or D'yakonov-Perel' (DP) \cite{dyakonov_spin_1972} spin relaxation model. In the EY case, the spin diffusion length is a function of the charge conductivity: $\ls(\sigma(d_\lin{NM})) = (\sigma(d_\lin{NM})/\sigma_\lin{bulk})\ls^\lin{max}$. The thickness-dependent conductivity and bulk conductivity $\sigma_\lin{bulk}$ are both determined by four-probe resistance measurements (see Section \ref{sec:t_depR}). By contrast, for DP spin relaxation, $\ls$ is independent of charge conductivity \cite{boone_spin-scattering_2015}. Additionally, the spin Hall conductivity in Eq. \eqref{eq:StilesHaney}  can be attributed to intrinsic or extrinsic SOC. For intrinsic spin Hall, $\sigmaSH$ is independent of charge conductivity, while for extrinsic SHE, $\sigmaSH(d_\lin{NM}) = \thetaSH \sigma(d_\lin{NM})$, where $\thetaSH$ is fixed. Fits using the four combinations of these models are shown in Fig. \ref{fig:sigmaSOTandAlpha}, with results collected in Table \ref{tab:SpinParams}. To distinguish between the quality of fit for the different models, we utilize a $\chi^2$ test.

%\subsection{$\chi^2$ Analysis for Quality of Fit}

The $\chi^2$ values for each fit of the SOT and damping data is calculated as $\chi^2 \equiv \sum_i^n (y_i - f_i)^2/\sigma_i^2$, where $y_i$ is the measured value, $f_i$ is the calculated value based on the fit model, and $\sigma_i^2$ is the measured variance, for each of $n$ measurements. Results are shown in Table \ref{tab:chi-sq}. Using the cumulative distribution function (CDF) of a $\chi^2$ distribution for each fit, with $\nu = n - p$ degrees of freedom, and $p$ fit parameters, we also calculate the joint probability with which we can reject the null hypothesis in which there is no relationship between our measurements and the given model. The CDF is determined by

\begin{equation}
\mbox{CDF}(\chi^2) = \int\limits_0^{\chi^2} \frac{t^{\nu/2-1} e^{-t/2}}{\Gamma(\nu/2) 2^{\nu/2}} dt
\end{equation}

\noindent where $\Gamma(x) = (x-1)!$. The joint probability is calculated as the product of $(1 - \mbox{CDF}(\chi^2))$ for the two fits. The EY/extrinsic model provides the highest confidence that we can reject the null hypothesis. Because this analysis reveals EY spin relaxation with extrinsic SHE as the best fit to our data, we focus on the fitted parameters from that model combination in the discussion below.

\begin{table*}
\begin{centering}
\begin{tabular}{l | c | c | c }
\toprule

Fit model & $\chi^2$ (SOT) & $\chi^2$ (damping) & Joint Probability \\
\hline

EY + ext & 0.668 & 3.696 & 0.89 \\
EY + int & 3.123 & 12.032 & 0.11 \\
DP + ext & 8.149 & 5.715 & 0.07 \\
DP + int & 8.819 & 6.101 & 0.05 \\
%data from ''Simultaneous Damping and SOT iterative fit results Python v2 w_cumulative probability.xlsx'' in Processed Data/10-2017
\botrule
\end{tabular}
\caption{$\chi^2$ values for SOT fit (Fig. \ref{fig:sigmaSOTandAlpha}(a)) and simultaneous damping fit (Figs. \ref{fig:sigmaSOTandAlpha}(b) and \ref{fig:alpha_dPy}). The joint probability represents the confidence with which we can reject the null hypothesis.}
\label{tab:chi-sq}
\end{centering}
\end{table*}
 
By choosing to enforce reciprocity, we find that the fraction of spin current absorbed by the Pt layer (which produces the damping-like AC charge currents) reaches a maximum of $(37 \pm 6)$\% for the thickest Pt layers. This is comparable to previous findings of large SML at Co/Pt interfaces \cite{nguyen_spin-flipping_2014} and Pt/Cu interfaces \cite{kurt_spin-memory_2002}. The different contributions to the total measured damping are represented as shaded areas in Fig. \ref{fig:sigmaSOTandAlpha}(b), with a color code to match Fig. \ref{fig:SpinPumping_CurrentDivider}. Note that only the contribution from spin pumping into Pt is Pt-thickness dependent. The self-consistent fit also results in a spin diffusion of length of $\lambda_\lin{s,Pt}^\lin{max} =$ \SI{4.2(1)}{\nm}, $\Gud = $\SI{1.3(2)e15}{\ohm^{-1} \m^{-2}}, which is in good agreement with the maximum theoretical value for Pt of $\Gud = $ \SI{1.07e15}{\ohm^{-1} \m^{-2}} \cite{liu_interface_2014}, given the estimated error, and $\sigmaSH^\lin{bulk} = $ \SI{2.36(4)e6}{\ohm^{-1}\m^{-1}}. This corresponds to a spin Hall angle of $0.387 \pm 0.008$. While this $\thetaSH$ is among the largest reported for Pt \cite{zhang_role_2015, pai_dependence_2015}, it is a necessary logical conclusion that with less spin current driven into the NM (on account of SML), a larger spin-to-charge conversion efficiency is required to fit the data than would be otherwise obtained if the SML were negligible. We furthermore stress that the phenomenological value for $\sigmaDL^\lin{SOT}$ (the asymptotic value in Fig. \ref{fig:sigmaSOTandAlpha}(a)) is comparable to that measured with other techniques (\SI{5.8e5}{\ohm^{-1}\m^{-1}} for AlO$_x$(2)\slash{}Co(0.6)\slash{}Pt(3) \cite{garello_symmetry_2013}, \SI{4.8e5}{\ohm^{-1}\m^{-1}} for Ta(2)\slash{}Pt(4)\slash{}Co$_{50}$Fe$_{50}$(0.5)\slash{}MgO(2)\slash{}Ta(1) \cite{pai_dependence_2015}, and $\approx$\SI{2.5e5}{\ohm^{-1}\m^{-1}} for Ta(1)\slash{}Pt($d_\lin{Pt}$)\slash{}Co(1)\slash{}MgO(2)\slash{}Ta(1) \cite{nguyen_spin_2016}). This indicates consistency of the SOC strength of the Pt layers in each of these experiments, and stresses the importance of characterizing spin loss mechanisms to optimize SOT for magnetic switching.

Our finding that the data are best fit with an extrinsic SHE model is somewhat surprising, given that it conflicts with some previous experimental work \cite{nguyen_spin_2016} and theoretical expectations \cite{guo_intrinsic_2008}. Qualitatively, both intrinsic and extrinsic SHE models are seen to describe the data quite well, given that the fit parameters can adjust to compensate for differences in the models, as is seen by the various fits in Fig. \ref{fig:sigmaSOTandAlpha}(a). Nevertheless, the $\chi^2$ analysis makes a clear distinction. Finally, the value for $\sigmaSH$ determined here is more than 5 times larger than the \SI{0}{\kelvin} prediction by Guo et al. (using their result of $\sigma_{xy} = 2.2 \times 10^5 (\hbar/e)$ \SI{}{\ohm^{-1} m^{-1}}, and setting $\sigmaSH = 2 \sigma_{xy}$ to account for the total spin current due to both up and down spins \cite{guo_intrinsic_2008}). This implies that the extrinsic effect dominates in our sputtered thin film systems where interfaces and crystal defects likely play a major role in determining the spin-orbit physics \cite{amin_spin_2016-1}. 

It is possible that some amount of intrinsic SHE is present in addition to the extrinsic effect, as discussed by Sagasta, et al. \cite{sagasta_tuning_2016}. In that work, the authors show that the total effective spin Hall conductivity $\sigmaSH^\lin{eff}$ can be described by:

\begin{equation}
\sigmaSH^\lin{eff} = \sigmaSH^\lin{int} + \thetaSH \sigma_\lin{Pt}
\label{eq:sigmaSH_combined}
\end{equation}

\noindent where $\sigmaSH^\lin{int}$ is the intrinsic spin Hall conductivity, and the second term describes the extrinsic effect as we have modeled it here. The Pt conductivities studied here (from $\approx$\SI{3e6}{\ohm^{-1}\meter^{-1}} to \SI{6e6}{\ohm^{-1}\meter^{-1}}) fall within the transition from intrinsic- to extrinsic-regimes, as described in Ref. \citenum{sagasta_tuning_2016}. Therefore, depending on the details of the spin and momentum scattering that govern $\thetaSH$, the extrinsic term in Eq. \eqref{eq:sigmaSH_combined} can easily be the dominant effect. Furthermore, we see no evidence of a large interfacial source of spin Hall conductivity, as in Ref. \citenum{wang_giant_2016}, which would manifest as a non-zero intercept of $\sigmaDL^\lin{SOT}$ in the limit of $d_\lin{Pt} \rightarrow 0$.

\begin{table*}
\begin{centering}
\begin{tabular}{l | c | c | c | c | c }
\toprule

Fit model & $\Gud \left(\times 10^{14}\right.$ \SI{}{\ohm^{-1}\m^{-2}}) & $\epsilon = \dfrac{\alpha_\lin{sp,Pt}}{(\alpha_\lin{sp,Pt} + \alpha_\lin{SML})}$ & $\ls$ (nm) & $\sigmaSH \left(\times 10^{6}\right.$ \SI{}{\ohm^{-1}\m^{-1}})& $\thetaSH = \dfrac{\sigmaSH}{\sigma_\lin{Pt}}$  \\
\hline

EY + ext & \SI{13(2)}{} & \SI{0.37(6)} &\SI{4.2(1)}{} & \SI{2.36(4)}{} & \SI{0.387(8)}{} \\
EY + int & \SI{5.6(1)}{} & \SI{0.20(4)} & \SI{6.7(3)}{} & \SI{5.3(3)}{} & \SI{0.86(5)}{} \\
DP + ext & \SI{3.0(3)}{} & \SI{0.19(2)} & \SI{2.5(2)}{} & \SI{11.6(4)}{} & \SI{1.91(6)}{} \\
DP + int & \SI{2.2(3)}{} & \SI{0.13(2)} & \SI{3.7(3)}{} & \SI{13.5(5)}{} & \SI{2.22(8)}{} \\
%data from ''Simultaneous Damping and SOT iterative fit results Python v2 w_cumulative probability.xlsx'' in Processed Data/10-2017
\botrule
\end{tabular}
\caption{Comparison of fitted values for $\Gud$, $\epsilon$, $\ls$, $\sigmaSH$, and $\thetaSH$ using different models for the source of spin relaxation (EY or DP) and SHE (intrinsic or extrinsic). For EY models, the spin diffusion length is reported as $\lambda_\lin{s}^\lin{max}$. }
\label{tab:SpinParams}
\end{centering}
\end{table*}

\subsection{Isolating the normal-metal layer contribution to sample inductance}
\label{sec:ControlSamples}
\begin{figure}
	\centering
	\includegraphics[width=\linewidth]{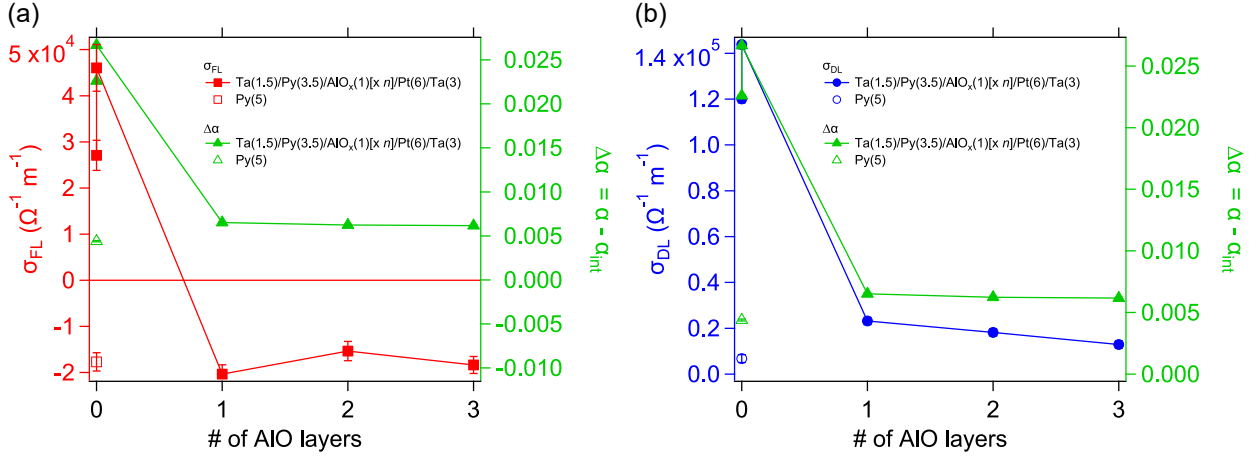}
	\caption{(a) FL  and (b) DL conductivities for samples with AlO$_x$ $[\times n]$ (where $n$ = 1, 2, or 3) blocking layers inserted between Py(3.5) and Pt(6). Also shown is a sample in which Py(5) is deposited directly onto SiO$_2$ (open symbols). Note that Py(3.5)/Py(6) (direct contact) was re-grown and re-measured as a part of the AlO$_x$ series (duplicate data points for zero AlO$_x$ repeats). The lower data point for both $\sigmaFL$ and $\sigmaDL$ at zero AlO$_x$ layers is that from the main text.}
\label{fig:sigma_controls}
\end{figure}

To better understand the influence of the normal metal layers (Ta seed, Pt or Cu spin sink, and Ta cap) on the perturbative inductance---and hence, the extracted FL and DL conductivities---that the sample contributes to a VNA-FMR measurement, we measured several control samples. First, we inserted an AlO$_x$ layer between the Py and the Pt in order to block spin pumping into the Pt \cite{zink_efficient_2016}. To do so, \SI{1}{\nm} of Al was sputter deposited onto the Py and subsequently oxidized for 10 minutes under \SI{5}{\torr} of O$_2$. The AlO$_x$ layer deposition and oxidation steps were repeated 1, 2, or 3 times, to ensure complete blocking of spin pumping. As can be seen in Fig. \ref{fig:sigma_controls}(b), the AlO$_x$ layers effectively reduce the damping by blocking spin pumping. This reduction correlates strongly with a reduction in $\sigmaDL$, confirming its signature as the damping-like conductivity. 

By contrast, $\sigmaFL$ actually changes sign with the introduction of the AlO$_x$ layers (Fig. \ref{fig:sigma_controls}(a)). The contribution to $\sigmaFL$ by Faraday-type pickup in the Pt cannot be eliminated by the AlO$_x$ barrier, since the Pt can still inductively couple to the precessing magnetization in the Py. The Faraday contribution clearly adds a negative contribution to $\sigmaFL$, as $\sigmaFL$ becomes increasingly negative with thicker Pt and Cu layers, as in Fig. \ref{fig:sigmaFL_DL}(a). Therefore, the AlO$_x$ barrier might eliminate the $\sigmaFL^\lin{SOT}$ contribution at the top Py interface. Nevertheless, even for Py deposited directly on SiO$_2$ (open square symbol), there remains a negative total $\sigmaFL$, perhaps due to the interface asymmetry that remains between the top and bottom Py interfaces. 

The control samples also elucidate the impact of the Ta layers on our measurements. We note that Eq. \eqref{eq:Geff} does not explicitly include SML at the Ta interface. Using the data from Fig. \ref{fig:sigma_controls}(b), we find that this simplification is justified. For these samples, we measured a total damping of $\alpha_\lin{tot} = 0.0104 \pm 0.0002$. If we set $\Gud_\lin{Py/Ta} = $ \SI{7.4e14}{\ohm^{-1}\m^{-2}} (the Sharvin value for Ta \cite{liu_interface_2014}), and use our measured conductivity of $\sigma_\lin{Ta} =$ \SI{8.91(2)e5}{\ohm^{-1}\m^{-1}}, we obtain $\alpha_\lin{sp,Ta} = 0.004$ (the amount depicted in Fig. \ref{fig:SpinPumping_CurrentDivider}). Therefore, when damping pathways into the Pt are blocked, the intrinsic damping plus spin pumping into the Ta accounts for all but 0.0023 of the total damping. Assigning this small amount of excess damping to SML at the Ta interface would reduce the contribution of SML at the Pt interface by less than 20\% and the values for spin Hall conductivity and spin Hall angle in Pt by only 10\%. 

Finally, we fabricated samples without any seed or capping layers. For Py(5) deposited directly onto SiO$_2$, $\sigmaDL$ is only 5\% of its value for Py(3.5)/Pt(6) (see open circle data point in Fig. \ref{fig:sigma_controls}(b)). The residual damping (beyond the intrinsic value) and damping-like conductivity for this sample could stem from the oxidized top surface or interfacial asymmetries, as well as less-than-optimal Py crystal structure, since no Ta seed layer was used.

In the cases of both $\sigmaFL$ and $\sigmaDL$, some residual signal remains even when spin pumping into the Pt is effectively blocked, or the seed and capping layers are eliminated entirely. Therefore, it is not surprising that even for our control samples in which Pt is replaced with Cu (with its weak spin-orbit interaction), some weak sources of spin-to-charge conversion (interfacial or otherwise) persist.

%From ``AlO thickness series (revised).pxp''

\section{Conclusion}
To summarize, by use of simultaneously acquired damping and iSOT data, we are able to properly assign the portions of damping enhancement incurred by a FM/NM bilayer due to the parallel channels of SML and spin pumping into the NM. These results suggest that Pt is indeed a promising material for spintronic applications. Our data also validate previous suggestions that interface engineering will be crucial for the optimization of SOT in multilayer systems \cite{nguyen_spin-flipping_2014, rojas-sanchez_spin_2014, pai_dependence_2015, zhang_role_2015}.

\section{Appendix}
\subsection{Pt thickness-dependent resistivity}
\label{sec:t_depR}
\begin{figure}
	\centering
	\includegraphics[width=0.5\linewidth]{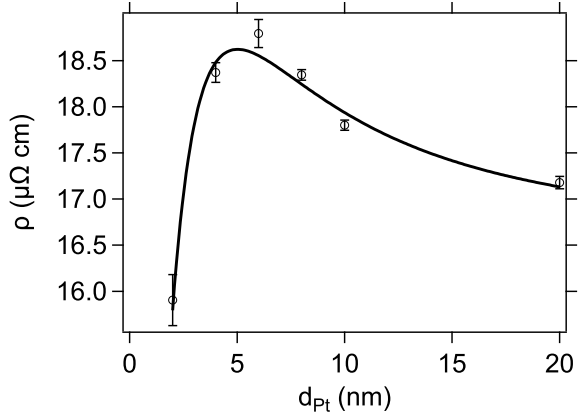}
	\caption{Thickness-dependent resistivity, measured for subtstrate/Ta(1.5)/Py(3.5)/Pt($d_\lin{Pt}$)/Ta(3) as a function of Pt thickness.}
\label{fig:RPt_dPt}
\end{figure}

To extract the Pt contribution to the total measured stack resistance, we have developed a model for the metallic multilayer stack to account for different conductivities in the bulk and at the metal interfaces. In this model, the interfacial conductivity $\sigma_\lin{int}$ at the Py/Pt interfaces decays exponentially to the Pt bulk value, $1/\rho_0$, with increasing distance from the interface.  Position-dependent conductivity through the Pt thickness can therefore be approximated as the sum of bulk and interfacial contributions:

\begin{equation}
\sigma(z) = \frac{1}{\rho_0}\left[1 - \exp\left(\dfrac{-z}{\sigma_\lin{int} \rho_0 \lambda}\right)\right] + \sigma_\lin{int}\exp\left(\dfrac{-z}{\sigma_\lin{int} \rho_0 \lambda}\right)
\end{equation}

\noindent where $\rho_0$ is the bulk resistivity, $\sigma_\lin{int}$ is the interfacial conductivity, and $\lambda$ is the bulk mean free path. The length scale $\sigma_\lin{int} \rho_0 \lambda$ describes the effective thickness over which the conductivity is determined by $\sigma_\lin{int}$. When $\sigma(z)$ is integrated over the Pt thickness from $z=0$ to $z=d_\lin{Pt}$, we obtain a final result for thickness-dependent resistivity:

\begin{equation}
\rho(d_\lin{Pt}) = \frac{\rho_0}{\left[1 + \left(\dfrac{\sigma_\lin{int} \rho_0 \lambda}{d_\lin{Pt}}\right)(\rho_0 \sigma_\lin{int} - 1) \left[1 -  \exp\left(\dfrac{-d_\lin{Pt}}{\sigma_\lin{int} \rho_0 \lambda}\right)\right] + \left(\dfrac{1}{R_\lin{other}}\right) \left(\dfrac{\rho_0}{d_\lin{Pt}}\right) \right]}
\end{equation}

\noindent where $R_\lin{other}$ represents the sheet resistances of any fixed-thickness metallic layers (here, Py and Ta). We use a calculated mean free path for our samples (\SI{12.79}{\nm}) by scaling a literature value \cite{fischer_mean_1980} (\SI{13}{\nm}) by the ratio of our measured bulk resistivity to the literature value for bulk resistivity. From the fit in Fit. \ref{fig:RPt_dPt}, we obtain $\sigma_\lin{int} = $ \SI{1.29(9)e6}{\ohm^{-1}\m^{-1}}, $\rho_0 = $ \SI{1.63(1)e-7}{\ohm\m}, and $R_\lin{other} = $ \SI{138(3)}{\ohm}. These values are used to obtain the thickness-dependent conductivity of the Pt layer, required in Eqs. \eqref{eq:Geff} and \eqref{eq:StilesHaney}.

%plotted in PyPt PtTS, interfacial resistivity fitting.pxp

%\bibliographystyle{apsrev4-1}
\bibliography{Physics}

\end{document}